\documentclass[a4paper,review,11pt,authoryear]{elsarticle}
\usepackage{amsmath,bm,hyperref,url,fullpage,mathtools,booktabs,bbm}
\usepackage{paralist,enumitem,todonotes,natbib}
\usepackage{multirow}
\usepackage{threeparttable}
\usepackage{subfig}

\usepackage{caption}
\providecommand{\tightlist}{%
  \setlength{\itemsep}{0pt}\setlength{\parskip}{0pt}}

\usepackage[a4paper, margin=1in]{geometry}

\usepackage{algorithm}
\usepackage{algorithmicx}
\usepackage{algpseudocode}

\usepackage{verbatim}

\DeclareMathOperator*{\argmin}{arg\,min}

\usepackage{hyperref}
\hypersetup{
  colorlinks=true,
  linkcolor=blue,
  citecolor=blue,
  urlcolor=blue}

\let\pkg=\texttt
\let\proglang=\textsf
\let\method=\textsf

\journal{arXiv}
\begin{document}
\date{}
\begin{frontmatter}

  \title{Forecast with Forecasts: Diversity Matters}
  \author[mainaddress]{Yanfei Kang} \ead{yanfeikang@buaa.edu.cn} \ead[url]{https://orcid.org/0000-0001-8769-6650}
  \author[mainaddress]{Wei Cao} \ead{caowei08@buaa.edu.cn}
  \author[secondaryaddress]{Fotios Petropoulos} \ead{f.petropoulos@bath.ac.uk} \ead[url]{https://orcid.org/0000-0003-3039-4955}
  \author[thirdaryaddress]{Feng Li\corref{cor}} \ead{feng.li@cufe.edu.cn} \ead[url]{https://orcid.org/0000-0002-4248-9778}

  \cortext[cor]{Corresponding author}
  \address[mainaddress]{School of Economics and Management, Beihang
    University, Beijing 100191, China.}
  \address[secondaryaddress]{School of Management, University of Bath, Claverton Down, Bath BA2 7AY, UK.}
  \address[thirdaryaddress]{School of Statistics and Mathematics, Central University of Finance and Economics, Beijing 102206, China.}

  \begin{abstract}
     Forecast combinations have been widely applied in the last few decades to improve forecasting. Estimating optimal weights that can outperform simple averages is not always an easy task. In recent years, the idea of using time series features for forecast combination has flourished. Although this idea has been proved to be beneficial in several forecasting competitions, it may not be practical in many situations. For example, the task of selecting appropriate features to build forecasting models is often challenging. Even if there was an acceptable way to define the features, existing features are estimated based on the historical patterns, which are likely to change in the future. Other times, the estimation of the features is infeasible due to limited historical data. In this work, we suggest a change of focus from the historical data to the produced forecasts to extract features. We use out-of-sample forecasts to obtain weights for forecast combinations by amplifying the diversity of the pool of methods being combined. A rich set of time series is used to evaluate the performance of the proposed method. Experimental results show that our diversity-based forecast combination framework not only simplifies the modelling process but also achieves superior forecasting performance in terms of both point forecasts and prediction intervals. The value of our proposition lies on its simplicity, transparency, and computational efficiency, elements that are important from both an optimisation and a decision analysis perspective.
  \end{abstract}

  \begin{keyword}
    Forecasting \sep Forecast Combination \sep Forecast Diversity \sep Prediction Intervals \sep Empirical Evaluation
  \end{keyword}

\end{frontmatter}

\section{Introduction}

\label{sec:introduction}

Many real-world problems are too complex for a single model that assumes a specific data generation process. Dating
back to 1818, Laplace stated that ``In combining the results of these two methods, one can
obtain a result whose probability law of error will be more rapidly decreasing''
\citep{CLEMEN1989559}. The literature shows that model combinations improve
overall performance in a variety of research areas, such as
regression \citep[e.g.,][]{mendes2012ensemble}, classification
\citep[e.g.,][]{rokach2010ensemble}, anomaly detection
\citep[e.g.,][]{perdisci2006using}, and time series forecasting
\citep[e.g.,][]{de2000review}. A recent overview of forecast combinations is provided in Section 2.5 of the encyclopedic review article by \cite{Petropoulos2020-ec}.

The motivation for forecast combinations has focused on finding the optimal weights of combining different
\emph{forecasts}, which are values at certain specific future times based on multiple
forecasting methods. The seminal work of \cite{bates1969combination} in the area of combining
forecasts suggests that forecast combinations can improve forecasting accuracy, provided that the sets
of forecasts contain some independent information. The usefulness of forecast combinations
has been demonstrated since then by numerous researchers using a variety of weighting
methods \citep[e.g.,
][]{winkler1983combination,mostaghimi1996combining,Watson2004Combination,fotios2015,montero2018fforma,kang2019gratis}.

Despite the large number of studies on forecast combinations, the ``forecast combination puzzle''
--the arithmetic mean performing better than more sophisticated combination methods in
some applications-- remains hard to tackle~\citep{Watson2004Combination,smith2009simple,claeskens2016forecast,Petropoulos2020-uf}. This situation is presumably related to that of equally weighted models in linear prediction~\citep{dawes1979robust}.
We summarize the main reasons for the  forecast combination puzzle as follows. On the one hand, the
optimal weights estimated by forecast combination are often sensitive to historical
data, and thus it is difficult to assemble robust forecasts that could consistently outperform a simple average.
On the other hand, the merits of forecast combinations stem from
independent information across multiple forecasts, which is further explained in Section
\ref{sec:method}. If all forecasts are close to identical, forecast combinations will be
close to a simple average. Therefore, an optimal forecast
combination depends to some degree on the \emph{diversity} of the individual forecasts, which is nonetheless
difficult to satisfy in reality \citep{thomson2019combining}.

Because the relative performance of different forecast methods changes,
depending on the nature of the time series \citep{reid1972comparison}, one way
to forecast a time series involves feature-based forecasting. The majority of the studies
in this line of work focus on developing rules or selecting the best
forecast model or averaging the models according to the historical features of the
data~\citep[e,g,][]{collopy1992rule-based,meade2000evidence,WSH09,petropoulos2014horses}. A recent implementation of feature-based forecasting was
proposed by \citet{talagala2018meta}. They used 42 time series features to train a random forecast classifier to select the best forecasting method.
\citet{montero2018fforma} used the same set of features to estimate optimal combination weights through an algorithm named eXtreme Gradient Boosting \citep[XGBoost, ][]{Chen2016XGBoost}.
Their approach, \textsf{FFORMA} (Feature-based FORecast Model Averaging), placed second
in the M4 competition. Both of these approaches used meta-learning, meaning that a group (reference set) of series is used to model the links between the time series
and the out-of-sample performance of the available forecasting models. Then, given a new series and its features, the most suitable model is selected, or a set of weights
for a forecast combination is estimated.

Regardless of the task (model selection or model combination), a common challenge in
feature-based forecasting is the choice and estimation of time series features. Time
series features vary from tens to thousands \citep{fulcher2014highly,rtsfeatures}, and
choosing different sets of features will inevitably result in different forecasts and
varied performance.  Moreover, the studies reviewed above focus on extracting such
features by using the historical, observed data of each time series. For example, two
commonly used features are the strength of a trend and the strength of the
seasonality. The estimation of these two features is not unique, because of the existence
of several approaches that are typically based on different assumptions. However, even if
there was one acceptable way to define them, it would become inadequate because existing
features are estimated based on observed historical patterns that are likely to change
over time. Moreover, the estimation of some of the features might not be feasible or
robust in the case of a limited number of available past observations.  Finally, when the
chosen features involve large numbers, this might increase the computational time required
to estimate them.

In this paper, we suggest a change of the focus in producing forecasts from extraction of time series features from historical data. Specifically, we suggest
the use the out-of-sample forecasts from a pool of models and measure their \textit{diversity}, a feature that has been identified as a crucial factor
in improving the performance of forecast combinations \citep{thomson2019combining,Lichtendahl2020-bj}. Through meta-learning, we use a group of series to
model the diversity of their forecasts and the optimal combination weights by minimizing the total forecasting loss. Once the model has been trained,
and for any new series that needs to be forecast, we can calculate the combination weights based on the diversity of their forecasts produced by the models
in the pool and produce both point forecasts and prediction intervals. We empirically show that a single feature, the \textit{diversity} of the forecasts,
is sufficient to achieve levels of postsample performance similar to those of large set of features derived from estimates based on  historical data.

Our study is in line with other studies that exploit information from the forecasts without utilization of forecast diversity. For example, \cite{PetropoulosSiemsenREP}
proposed forecast representativeness and derived a new selection criterion that works remarkably well in cases of low signal to noise ratios in comparison with other established
selection criteria. The ability of the representativeness criterion to more often than not select the best (and avoid the worst) models leads to significant accuracy improvement both
in selecting single models and combinations across models. \citet{Zhao2020-fq} also used postsample forecasts as the input in a machine learning model as a step toward enhancing
the performance of standard statistical time series forecasting models. Although they also use forecasts as a feature, they do not explicitly focus on a specific aspect of the
forecasts (such as diversity), making it difficult to obtain insights into why their approach performs well. They also did not discuss how prediction intervals may be estimated.

Forecasting is vital for the efficient operation of supply chains \citep{Tliche2020-dd,Ali2017-lb} as well as other operations-related decisions. To support the efficacy of our proposition
for decision-making purposes, we offer not only large-scale empirical evaluations for the mean (point) forecasts but also for the uncertainty around this mean, in terms of quantile forecasts.
We also provide trade-off curves based on upper coverage levels versus upper prediction intervals that approximate utility functions related to inventory forecasting.

The rest of this paper is organised as follows. In Section~\ref{sec:method}, we describe the calculation of forecast diversity for forecast combinations and present a framework
of forecasts with forecasts. We demonstrates the superiority of the proposed approach via
extensive experiments in Section~\ref{sec:experiments}. Section~\ref{sec:discussion} provides our discussions and Section~\ref{sec:conclusion} concludes the paper.

\section{Forecast combination: diversity matters}
\label{sec:method}

\subsection{Diversity of forecasts}

Ambiguity Decomposition~\citep{krogh1995validation} indicates in the literature on machine learning that accuracy and diversity are two main factors that should be taken into consideration
when designing ensembles. In the forecasting community, many studies have emphasized the importance of the forecasting method's diversity pool when
constructing forecast combinations \citep{bates1969combination,batchelor1995forecaster,thomson2019combining}. \cite{Lichtendahl2020-bj}, in exploring why some combinations performed
better than others in the recent M4 competition~\citep{makridakis2018m4}, also identified diversity as an important factor for efficient forecast combinations, along with the robustness
of the individual models.

Ambiguity Decomposition can be easily applied to the forecast combination task. For a
given time series $\{y_t, t = 1, 2, \cdots, T\}$, we denote the $h$-th step forecast
produced by the $i$-th individual method as $f_{ih}$, where $i = 1, 2, \cdots, M$ and
$h = 1, 2, \cdots, H$. Furthermore, $M$ and $H$ are the number of algorithms in the
forecast pools and the forecast horizon, respectively. Let $f_{ch}$ be the $h$-th step
combined forecast given by $\sum_{i=1}^{M}w_if_{ih}$, where $w_i$ is the combination
weight for the $i$-th method. The overall mean squared error of a weighted forecast
combination model $MSE_{comb}$ over the whole forecast horizon $H$ can be written as
follows.
\begin{equation}
\begin{aligned}
MSE_{comb} & = \frac{1}{H} \sum_{i=1}^{H}\left( \sum_{i=1}^{M}w_if_{ih} - y_{T+h}\right)^2 \\
     & = \frac{1}{H}\sum_{i=1}^{H}\left[ \sum_{i=1}^{M}w_i(f_{ih} - y_{T+h})^2 - \sum_{i=1}^{M}w_i(f_{ih} - f_{ch})^2\right] \\
      & = \frac{1}{H}\sum_{i=1}^{H}\left[\sum_{i=1}^{M}w_i(f_{ih} - y_{T+h})^2 - \sum_{i=1}^{M-1} \sum_{j>i}^{M}w_iw_j(f_{ih}-f_{jh})^2\right] \\
      & = \sum_{i=1}^{M}w_i MSE_i - \sum_{i=1}^{M-1} \sum_{j>i}^{M}w_iw_jDiv_{i,j},
\end{aligned}
\label{eq:mse-comb}
\end{equation}
where $MSE_i$ represents the mean squared error for the $i$-th method. $Div_{i,j}$ denotes the degree of diversity between the $i$-th and $j$-th method in the forecast method pool,
which is defined as follows.
\begin{equation}
\label{eq:div}
   Div_{i,j} = \frac{1}{H}\sum_{i=1}^{H} (f_{ih}-f_{jh})^2.
\end{equation}

Equation~\eqref{eq:mse-comb} says that the mean squared error of the combined forecast is guaranteed to be less than or equal
to the weighted mean squared error of the individual forecasts. The second term in the last line of Equation~\eqref{eq:mse-comb} tells us how diverse the individual forecasts are.
Out of two combination methods with identical weighted mean squared error, the
one with greater diversity will have a lower overall squared error. That is, the more diversity existing in the forecast method pool leads to overall better forecasting accuracy.

\subsection{Diversity for forecast combination}

How can we exploit the diversity information among different forecasting methods for
forecast combination? As \citet{montero2018fforma} and
\citet{kang2019gratis} point out, we can use a group of series to estimate the forecast combination weights by linking a set
of time series features with the forecasting performances of the individual methods. The key point is
to find a set of features that can represent the information affecting forecasting
performance. We propose to use the pairwise diversity measures as a proper set of
features to represent the forecast diversity among different methods.

To make the diversity comparable between time series with different scales, we can scale the diversity measure in Equation~\eqref{eq:div} by averaging across all pairs of methods.
We use the scaled diversity in Equation~\eqref{eq:sdiv} for all the experiments in the following sections.
\begin{equation}
\label{eq:sdiv}
sDiv_{i,j} = \frac{\sum\limits_{h=1}^H(f_{ih}-f_{jh})^2}{\sum\limits_{i=1}^{M-1}\sum\limits_{j=i+1}^M\left[\sum\limits_{h=1}^H(f_{ih}-f_{jh})^2\right].}
\end{equation}

\begin{figure}
  \centering
  \includegraphics[width=\textwidth]{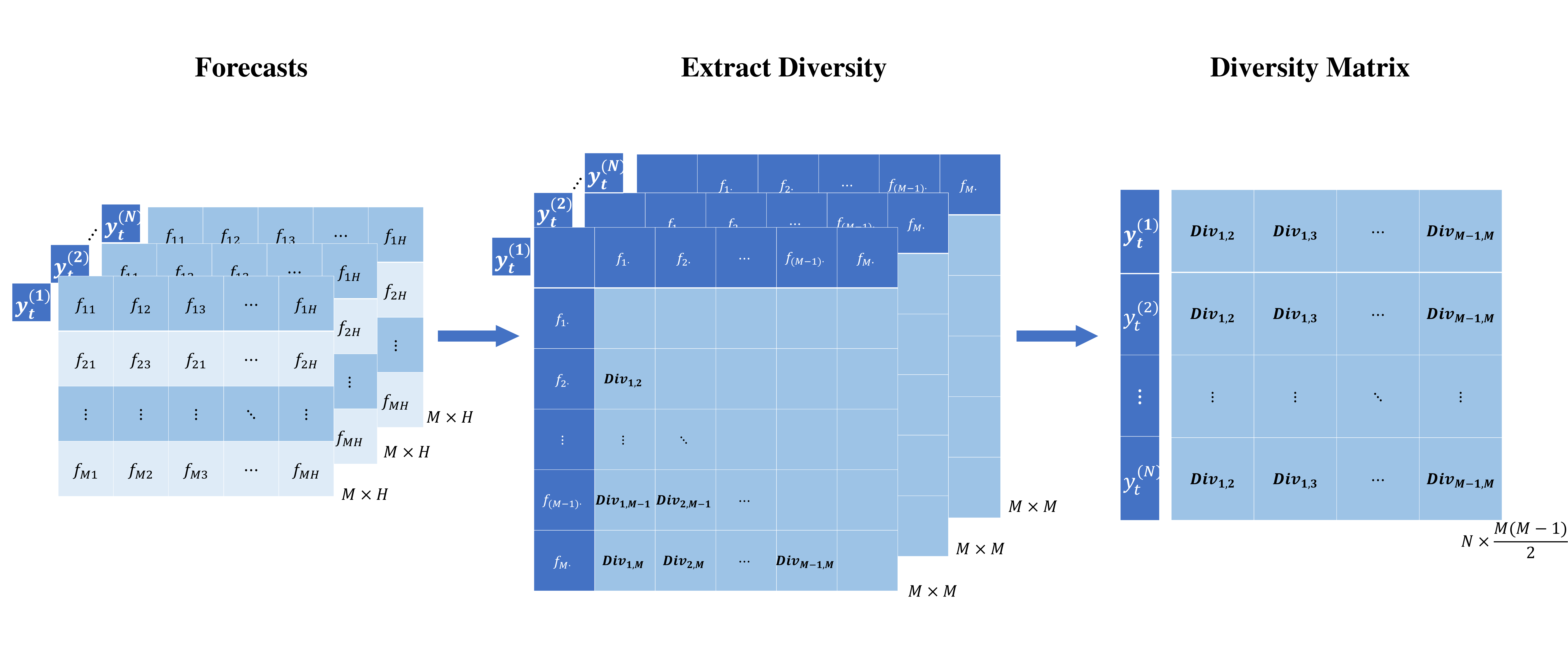}
  \caption{Diversity extraction from forecasts.}
  \label{fig:diversity}
\end{figure}

Figure~\ref{fig:diversity} shows the diversity extraction procedure in the context of
point forecasting. Given a time series data set $\{y_t^{(n)}\}_{n=1}^N$, for each time
series $y_t^{(n)}$, its $h$-th step point forecast produced by the $m$-th method is
denoted as $f^{(n)}_{mh}$, where $m = 1, 2, \cdots, M$ and $h = 1, 2, \cdots, H$.
Therefore, we can get an $M \times H$ matrix for each time series to forecast, thus,
representing the forecasts produced by the $M$ methods for the entire forecasting
horizon. Then the pairwise forecast diversity among the $M$ methods can be calculated by
using Equation~\eqref{eq:sdiv}. Thus, for each time series $y_t^{(n)}$, we get an
$M \times M$ symmetric matrix. We then concatenate the diversity measures in the lower
diagonal into a vector. This can be used as a feature vector for time series $y_t^{(n)}$
when estimating the corresponding forecast combination weights based on the feature-based
forecasting framework.  For a forecasting pool containing $M$ methods, we can construct
$M(M-1)/2$ pairwise diversity measures to extract the independent information in the pool.

The same procedure can be extended to the context of interval forecasting. For a forecasting pool containing $M$
methods, we can construct $M(M-1)/2$ pairwise diversity measures for both the upper and lower prediction intervals.

Assuming that we need to forecast $N$ time series, we obtain $N \times (M (M-1)/2)$ matrices (for the point forecasts and the upper and lower prediction intervals), where each row can
be viewed as a diversity (feature) vector for the corresponding series.

\subsection{Forecast with forecasts: the framework}\label{sec:framework}

To construct a forecast model using diversity, we tailor the state-of-the-art feature-based forecast model averaging (\textsf{FFORMA}) framework proposed by
\citet{montero2018fforma} to allow for the diversity of the forecasts as the inputs. Next, we
 estimate the combination weights based on the diversity information of the out-of-sample forecasts produced by the pool of methods being combined.

In the original \textsf{FFORMA} framework, estimations of forecast combination weights is
implemented by finding a function to assign weights to each forecasting method. To forecast with \textsf{FFORMA}, one first needs to extract $42$ features from the original time series and
calculate the overall weighted average (OWA) error of each forecast algorithm in the
forecast pool. Then, to obtain the optimal combination weights, features are linked with the OWA errors using the XGBoost algorithm, which
is an ensemble machine learning algorithm and can deal with regression or
classification problems by integrating plenty of decision tree models. We find that \textsf{FFORMA} has the following drawbacks: (1) a manual selection of features is required,
with specific features more appropriate than others in some applications and contextss, and (2) because \textsf{FFORMA} extracts features based on  historical data, it is not applicable
to time series with an inadequate length of historical data; in such cases the estimations of features may be unreliable.

\begin{figure}[!ht]
  \centering
  \includegraphics[width=1\linewidth]{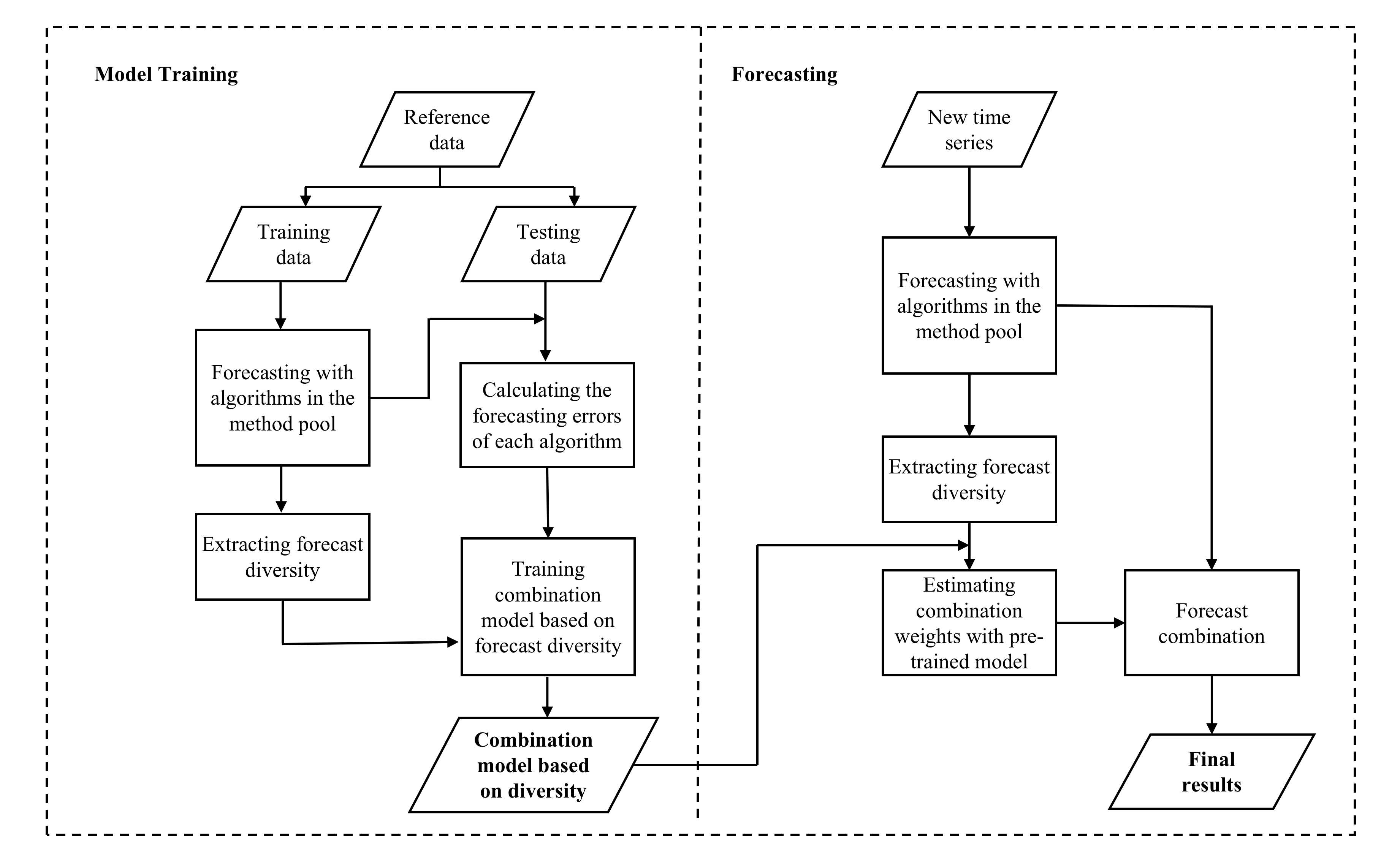}
  \caption{The framework of forecast with forecasts.}
  \label{fig:frame}
\end{figure}

To address the above problems, we propose a diversity-based forecast combination framework. It consists of two phases, namely model-training and
forecasting, as shown in Figure \ref{fig:frame}. In the model-training phase, we take all the series in a given data set (for which forecasts are required) as the reference data.
Each time series in the reference data is split into training and testing periods. The length of the testing period for each series is the same as its forecasting horizon.
We apply the forecasting methods in the pool by using the training periods, and extract the diversity matrix following Figure~\ref{fig:diversity} from the forecasts produced by
different forecasting methods on the testing periods. We then calculate the forecasting errors of each method and summarize them using an error metric. Finally, a forecast combination model
is trained, by minimizing the total forecasting loss, to estimate the combination weights for each series as a function of its forecast diversity. Once the model has been trained, weights
can be produced for any target series given the diversity of its forecasts produced by the method pool.

We present a detailed procedure forf constructing diversity-based forecasting combinations
in Algorithm \ref{alg:A}. Using the XGBoost algorithm, the following
optimization problem is solved to obtain the combination weights:
\begin{equation}
  \argmin_{w} \sum\limits_{n = 1}^N\sum_{i=1}^{M} w({Div_n})_i \times \text{Err}_{ni},
\end{equation}
where $Div_n$ indicates the forecast diversity of the $n$-th time series, $w({Div_n})_i$
is the combination weight assigned to method $i$ for the $n$-th time series based on the
diversity, and $\text{Err}_{ni}$ is the error produced by method $i$ for the
$n$-th time series. The combination weight $w({Div_n})_i$ is obtained by the output of
XGBoost model after a soft-max transformation:
\begin{equation*}
  w({Div_n})_i = \frac{\exp\{y(Div_n)_i\}}{\sum_{i=1}^M \exp\{y(Div_n)_i\}},
\end{equation*}
where $y(div_n)_i$ is the regression results from XGBoost. In other words, the key point of the combination model based on diversity is to search for the optimal weights that
can minimize the weighted error. After obtaining the combination weights, we can calculate the point and interval forecasts as below.
\begin{equation}
\begin{aligned}
    f_n = \frac{1}{M}\sum_{i=1}^{M}w_{ni}f_{ni},\\
    f_n^u = \frac{1}{M}\sum_{i=1}^{M}w_{ni}f_{ni}^u,\\
     f_n^l = \frac{1}{M}\sum_{i=1}^{M}w_{ni}f_{ni}^l,
\end{aligned}
\label{eq:forecasts}
\end{equation}
where $w_{ni}$ is the estimated weight for the $n$-th time series and the $i$-th method. And $f_{ni}$, $f_{ni}^u$, $f_{ni}^l$ are the point, upper, and lower forecast values of the $i$-th algorithm for $h$-th forecast step, respectively.

\begin{algorithm}[!ht]
  \caption{The framework of forecasts with forecasts.}
  \label{alg:A}
  \begin{algorithmic} [1]

    \Statex \textbf{Phase 1: Model training}
    \Require $y_{\text{ref}}$: a time series reference set; a forecasting pool consisting of $M$ methods.
    \Ensure
    Forecast combination model based on diversity.

    \For {$y_t^{(n)} \in y_{\text{ref}}$}
    \State Split time series $y_t^{(n)}$ into training and testing periods.
    \State Produce the forecasts using the $M$ methods.
    \State Extract the diversity vector: $Div_n$ (see Figure~\ref{fig:diversity}).
    \State Calculate the forecasting errors of each method in the pool on the testing data.
    \EndFor
    \State Estimate the combination model based on diversity with XGBoost, by minimizing the weighted errors:
    $$\argmin_{w} \sum\limits_{n = 1}^N\sum_{i=1}^{M} w({Div_n})_i \times \text{Err}_{ni}.$$ // \textit{Finish building combination model.}

    \Statex \textbf{Phase 2: Forecasting}
    \Require Pretrained model; $y_{\text{new}}$: a time series data set to be forecast.

    \Ensure Final forecasts of new time series $y_{new}$.

    \For {$y_t^{(m)} \in y_{\text{new}}$}
    \State Produce forecasts using the methods in the forecasting method pool.
    \State Extract the diversity vector: $Div_m$ (see Figure~\ref{fig:diversity}).
    \State Use the pretrained model to produce the optimal weight $w({Div_m})_i$ for method $i$.
    \State Combine the individual forecasts using $w({Div_m})_i$ and obtain the final forecasts.
    \EndFor  // \textit{Obtain final results.}
  \end{algorithmic}
\end{algorithm}

The merits of using diversity for forecast combinations are twofold. First, the process
of extracting diversity is straightforward and interpretable. The algorithm of
measuring the diversity between different methods involves a simple calculation, and
hence, it can reduce the computational complexity when extracting
features. Meanwhile, diversity-based forecast combinations require only the forecasts from
individual methods, avoiding exploiting information from other models as in
\citet{montero2018fforma}. Secondly, although traditional methods of time series feature extraction
\citep{fulcher2014highly,rtsfeatures,CHRIST201872} usually depend on the manual
choice of an appropriate set of features, our approach can be applied automatically without the need for expert knowledge and human interaction.

\section{Empirical evaluation}
\label{sec:experiments}

\subsection{Data}
We used the M4 data set \citep{makridakis2019} to evaluate the forecasting performance of the proposed
diversity-based forecast combination method in terms of both point and interval forecasting. The M4 data contains 100,000 time series with different seasonal
periods from different domains such as demographics,
finance, and industries. The lengths of the yearly, quarterly, monthly, weekly, daily, and hourly data lie
in the ranges of $[13, ~ 835]$, $[16, ~ 866]$, $[42, ~ 2794]$, $[80, ~ 2597]$,
$[93, ~ 9919]$ and $[700, ~ 960]$, respectively.  The corresponding forecasting horizons
are 6, 8, 18, 13, 14, and 48. The data set is publicly available in the \pkg{M4comp2018}
\proglang{R} package~\citep{montero2018m4comp2018}. We optimized the combination weights separately for each frequency by using the respective M4 series to form the reference data
(see Section~\ref{sec:framework} for more details).

\subsection{The forecasting pool of methods}\label{sec:pool}

Our forecasting pool consists of eight individual forecast methods; they are described in Table \ref{tab:fmethod}. Note that compared with the commonly used nine individual methods
in recent forecast combination studies \citep{kang2019bmsr,montero2018fforma,kang2020imaging,kang2019gratis}, we do not include the neural network time series forecasting method (nnetar)
because it does not produce prediction intervals. The eight forecasting methods in our pool are implemented in the \texttt{forecast} package in R \citep{hyndman2018forecast}.

\begin{table}[!ht]
  \centering
  \caption{The methods used for forecast combination. All these methods are implemented using the \pkg{forecast} \textbf{R} package.}
  \begin{tabular}{p{0.65\columnwidth}p{0.3\columnwidth}}
    \toprule
    Forecasting Method         & \textbf{R} implementation                 \\ \midrule
    \textbf{auto\_arima}: the ARIMA family of models~\citep{HK08}.             & \texttt{auto.arima() }          \\
    \textbf{ets}: the exponential smoothing state space family of  models~\citep{hyndman2002state}.      & \texttt{ets()}     \\
    \textbf{tbats}: the exponential smoothing state space model with a Box-Cox transformation, ARMA errors, trend and seasonal components ~\citep{de2011forecasting}.       & \texttt{tbats()}                          \\
    \textbf{stlm\_ar}: seasonal and trend decomposition using Loess with AR modeling  of the seasonally adjusted series. & \texttt{stlm()} with   \texttt{modelf = ar}                       \\
    \textbf{rw\_drift}: random walk with drift.                     & \texttt{rwf()}   with \texttt{drift=TRUE} \\
    \textbf{thetaf}: the theta method~\citep{assimakopoulos2000theta}.           & \texttt{thetaf()} \\
    \textbf{na\"ive}: the na\"ive   method.                                                  & \texttt{naive()} \\
    \textbf{sna\"ive}: the seasonal na\"ive method.                   & \texttt{snaive()}                         \\ \bottomrule
  \end{tabular}\label{tab:fmethod}
\end{table}

\subsection{Diversity extraction}\label{sec:divextraction}

Since we are producing both point and interval forecasts, we calculate the diversity features based on the upper and lower prediction intervals.  Considering the eight individual forecasting methods in the pool, we have 28 diversity features for upper prediction intervals, and 28 features for lower intervals. Therefore, in total, 56 diversity features are calculated for each time series. Note here we are not using the diversity of point forecasts, which are the midpoints of the prediction intervals for all the eight methods and do not contain more information than the upper and lower intervals.

\subsection{Forecasting evaluation metrics}\label{sec:metrics}

We use the mean absolute scaled error \citep[MASE,][]{hyndman2006another} to evaluate the point forecasts produced by our proposed combination model. MASE compares the forecast accuracy
between a specific forecast algorithm and the na\"ive method. It is defined as
 \begin{equation*}
   \text{MASE} = \frac{1}{H}\frac{\sum_{h=1}^{H}|f_h-y_{T+h}|}{\frac{1}{T-m}\sum_{t=m+1}^{T}|y_t-y_{t-m}|},
 \end{equation*}
where $H$ is the forecasting horizon, $T$ is the length of the historical data, $m$ is the frequency of the data, $y_{T+h}$ is the actual value of the time series at time $T+h$,
and $f_h$ is the forecast value at the $h$-th step.

To assess the performances of the generated prediction intervals, we use the mean scaled interval score~\citep[MSIS, ][]{gneiting2007strictly}, as used in the M4 competition
and other recent studies on forecast uncertainty estimation~\citep[e.g., ][]{spiliotis2020generalizing,kang2020deja}. The definition of MSIS is as follows.
\begin{equation*}
\mathrm{MSIS} = \frac{1}{H}\frac{\sum_{h=1}^{H}\left\{(U_h-L_h)+\frac{2}{\alpha}(L_h-y_{T+h})\mathbbm{1}\left\{ y_{T_h} < L_h\right\} + \frac{2}{\alpha}(y_{T+h} - U_h)\mathbbm{1}\left\{y_{T+h}>U_h\right\}\right\}}{\frac{1}{T-m}\sum_{t=m+1}^{T} \vert y_t-y_{t-m} \vert},
\end{equation*}
where $\left[ L_h, U_h \right]$ are the generated $(1-\alpha)100\%$ prediction intervals at the $h$-th step,  and $\mathbbm{1}$ is the indicator function, which equals to $1$ when
$y_{T+h}$ is within the postulated interval and returns $0$ otherwise.

In the training process of Algorithm~\ref{alg:A}, we apply a new cost function that takes both point and interval forecasting into consideration when optimizing the combination weights.
It is defined as
\begin{equation}
\label{eq:err}
    \text{Err} = \frac{1}{2}\left(\frac{\text{MASE}}{\text{MASE}_{\text{naive2}}}+\frac{\text{MSIS}}{\text{MSIS}_{\text{naive2}}}\right),
\end{equation}
where $\text{MASE}_{\text{naive2}}$ and $\text{MSIS}_{\text{naive2}}$ are the MASE and MSIS values of the naive2 forecasting method, derived from the na\"ive method on the seasonally
adjusted data~\citep{makridakis2018m4}. In this way, we are using the same group of features for both point and interval forecasts (see Section~\ref{sec:divextraction}) and the same cost
function for estimating the combination weights. The point and interval forecasts are then calculated following Equation~\eqref{eq:forecasts}.

\subsection{Point and interval forecasting performance}
\label{subsec:xgboost-results}

We compared the performance of point and interval forecasts of the proposed diversity-based forecast combination approach as shown in Algorithm~\ref{alg:A} against the \textsf{FFORMA}
approach~\citep{montero2018fforma} that uses XGBoost to link 42 statistical time series features with forecast errors. We also benchmarked against a simple average (\textsf{SA}) approach,
where the forecasts from all methods in the forecasting pool are combined with equal weights. Table \ref{tab:meanerrors} depicts the mean of the forecast errors across series from each
frequency. Entries in bold highlight that our method outperforms the \textsf{FFORMA} approach.
We can see that both our proposed method
(\textsf{Diversity}) and \textsf{FFORMA} outperform \textsf{SA}. Overall, the mean values of MASE and MSIS of the proposed approach are 18.71\% and 19.92\% lower, respectively,
compared with \textsf{SA}. In other words, the unequal weights produced by diversity-based modeling are effective and can help tackle the ``forecast combination puzzle.'' More importantly,
\textsf{Diversity}, without extracting and selecting sophisticated times series features, outperforms \textsf{FFORMA} when we focus on the mean MASE and MSIS values of the overall M4 data.
\textsf{Diversity} outperforms \textsf{FFORMA} in most frequencies of data, apart from the weekly and daily data.

\begin{table}[ht!]
  \centering
  \caption{Comparison of the mean MASE and MSIS values from our diversity-based forecast combination method
    (\textsf{Diversity}), \textsf{FFORMA} and forecast combination with simple averaging (\textsf{SA}). Entries in bold highlight that our method outperforms the FFORMA approach.}
  \label{tab:meanerrors}
\begin{tabular}{lrrrrrrrr}
    \toprule
    Method       & Overall       & Yearly       & Quarterly      & Monthly       & Weekly      & Daily       & Hourly          \\
    \midrule
    \multicolumn{8}{c}{MASE}                                                                                                   \\
    \textsf{SA}           & 1.9040        & 3.6907       & 1.2432         & 0.9813        & 6.3826            & 5.8921          & 3.3319          \\
    \textsf{FFORMA}       & 1.5586        & 3.0842       & 1.1220         & 0.8980        & \textbf{2.2309}   & 3.2464          & 0.8822          \\
    \textsf{Diversity}    & \textbf{1.5478}   & \textbf{3.0670}   & \textbf{1.1095}   & \textbf{0.8915} & 2.2744    & \textbf{3.2296}   & \textbf{0.8540} \\
    \textsf{FD}           & \textbf{1.5507}   & \textbf{3.0615}   & \textbf{1.1096}   & 0.8997          & 2.2639    & \textbf{3.2345}   & \textbf{0.8574} \\
    \midrule
    \multicolumn{8}{c}{MSIS}                                                                                                    \\
    \textsf{SA}           & 17.5077       & 42.0776       & 9.9248        & 8.3012         & 22.4778           & 31.5910        & 11.4214         \\
    \textsf{FFORMA}       & 14.5934       & 32.0185       & 9.2388        & 7.8189         & \textbf{16.0496}           & \textbf{27.7694}        & 6.6161          \\
    \textsf{Diversity}    & \textbf{14.0197}  & \textbf{30.3312}  & \textbf{8.7805}   & \textbf{7.6385}  & 16.4015   & 28.0220  & \textbf{6.3587} \\
    \textsf{FD}           & \textbf{14.0254}  & \textbf{30.3980}  & \textbf{8.7995}   & \textbf{7.6248}  & 16.0936   & 27.8723  & \textbf{6.3145} \\
    \bottomrule
\end{tabular}
\end{table}

To further verify the performance of \textsf{Diversity}, we  consider an approach that combines the features from \textsf{Diversity} and \textsf{FFORMA}. We combine the
42 statistical features and the 56 diversity features into a single set of time series
features and use them as the inputs into XGBoost to obtain the optimal weights for the nine
forecasting methods. The results for this approach, \textsf{FD} (\method{FFORMA + Diversity}), are also
represented in Table \ref{tab:meanerrors}. We observe that \textsf{FD} performs similarly with \textsf{Diversity}.

Note that in Table~\ref{tab:meanerrors}, \textsf{FFORMA}, \textsf{Diversity} and \textsf{FD} are using the same pool of forecasting methods and cost functions as described
in Section~\ref{sec:pool} and Section~\ref{sec:metrics} to make them comparable. Therefore, the results for  \textsf{FFORMA} are slightly different from those in \citet{montero2018fforma}
where a different pool (including NNETAR) and cost functions (OWA) are used.

\begin{figure}[!th]
\centering
\hspace{18mm}
\begin{minipage}{0.45\textwidth}
\includegraphics[width=\linewidth]{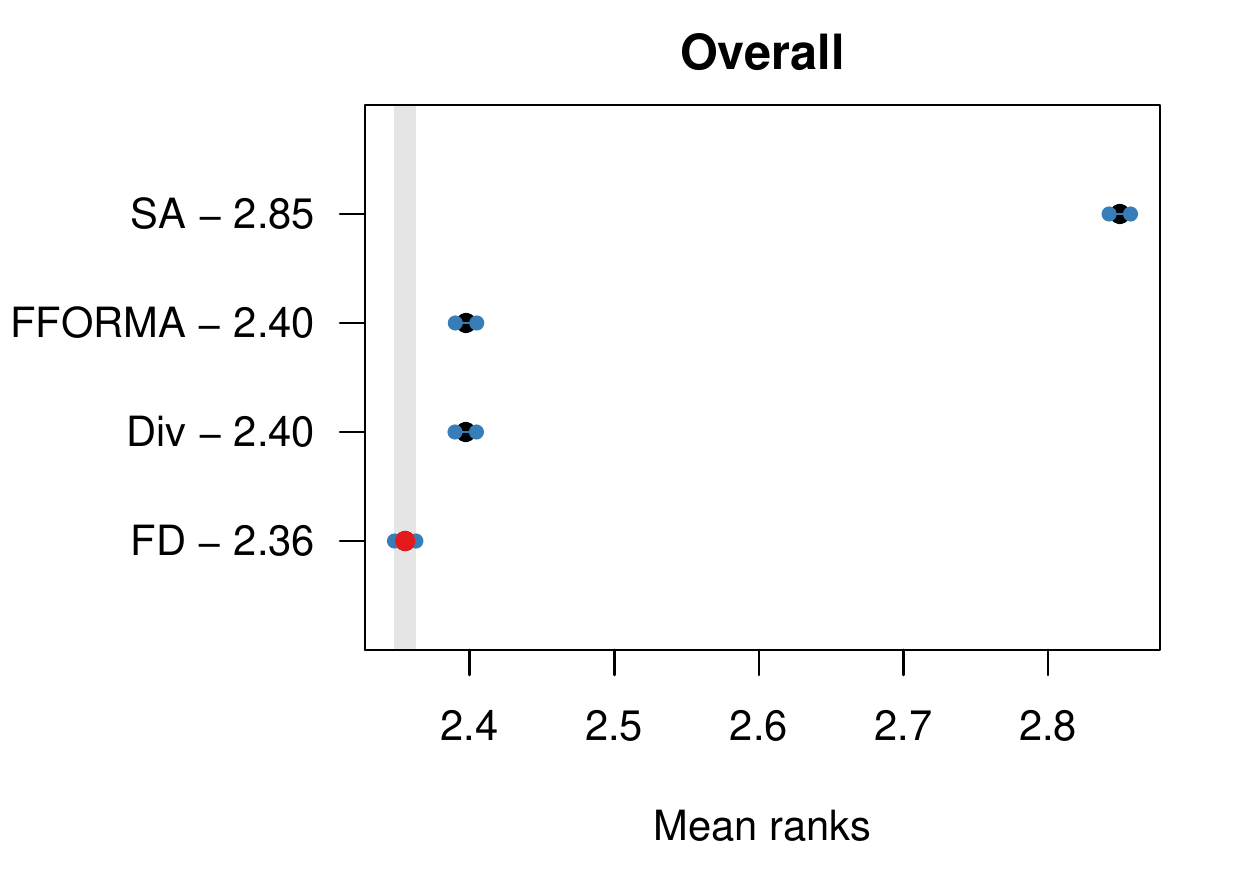}
\end{minipage}
\newline
\medskip
\begin{minipage}{0.45\textwidth}
\includegraphics[width=\linewidth]{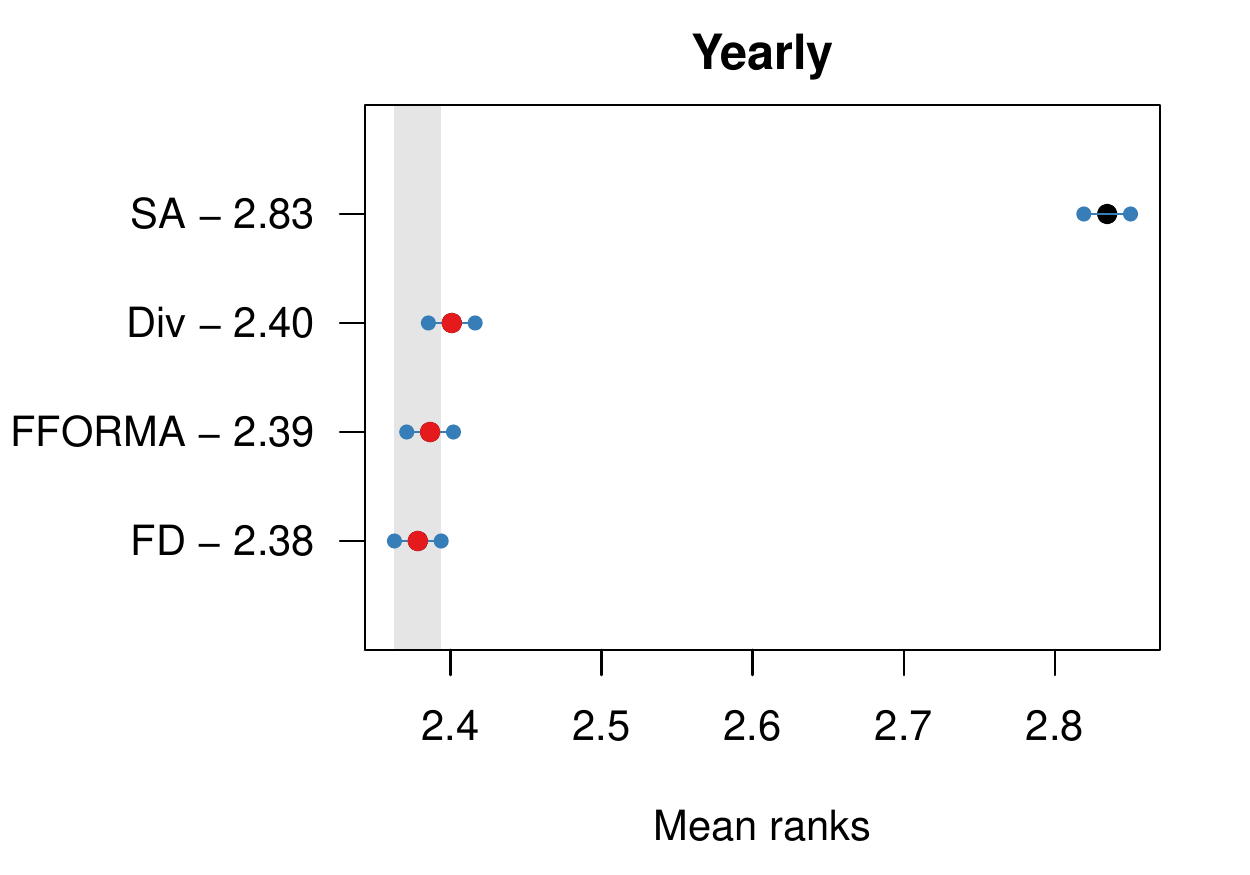}
\end{minipage}
\begin{minipage}{0.45\textwidth}
\includegraphics[width=\linewidth]{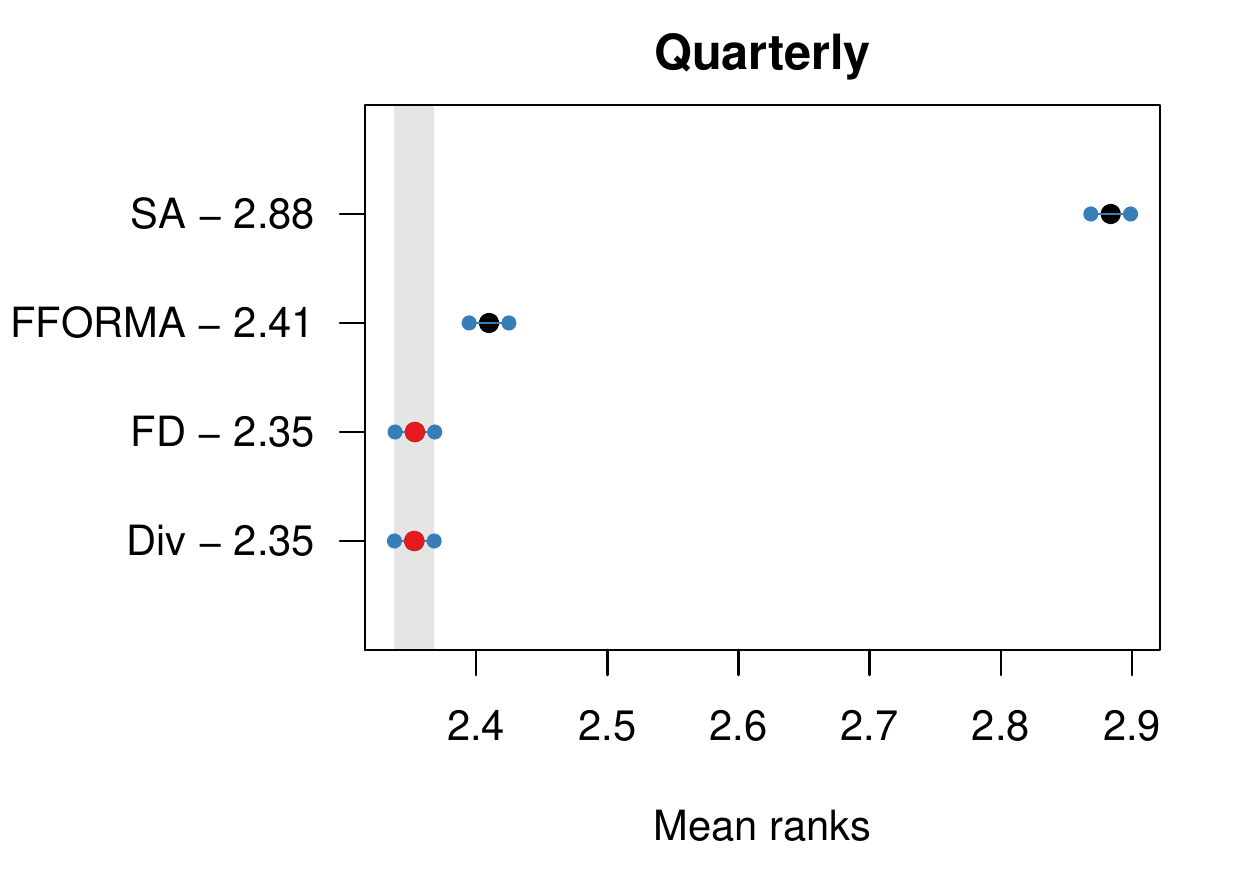}
\end{minipage}
\medskip
\begin{minipage}{0.45\textwidth}
\includegraphics[width=\linewidth]{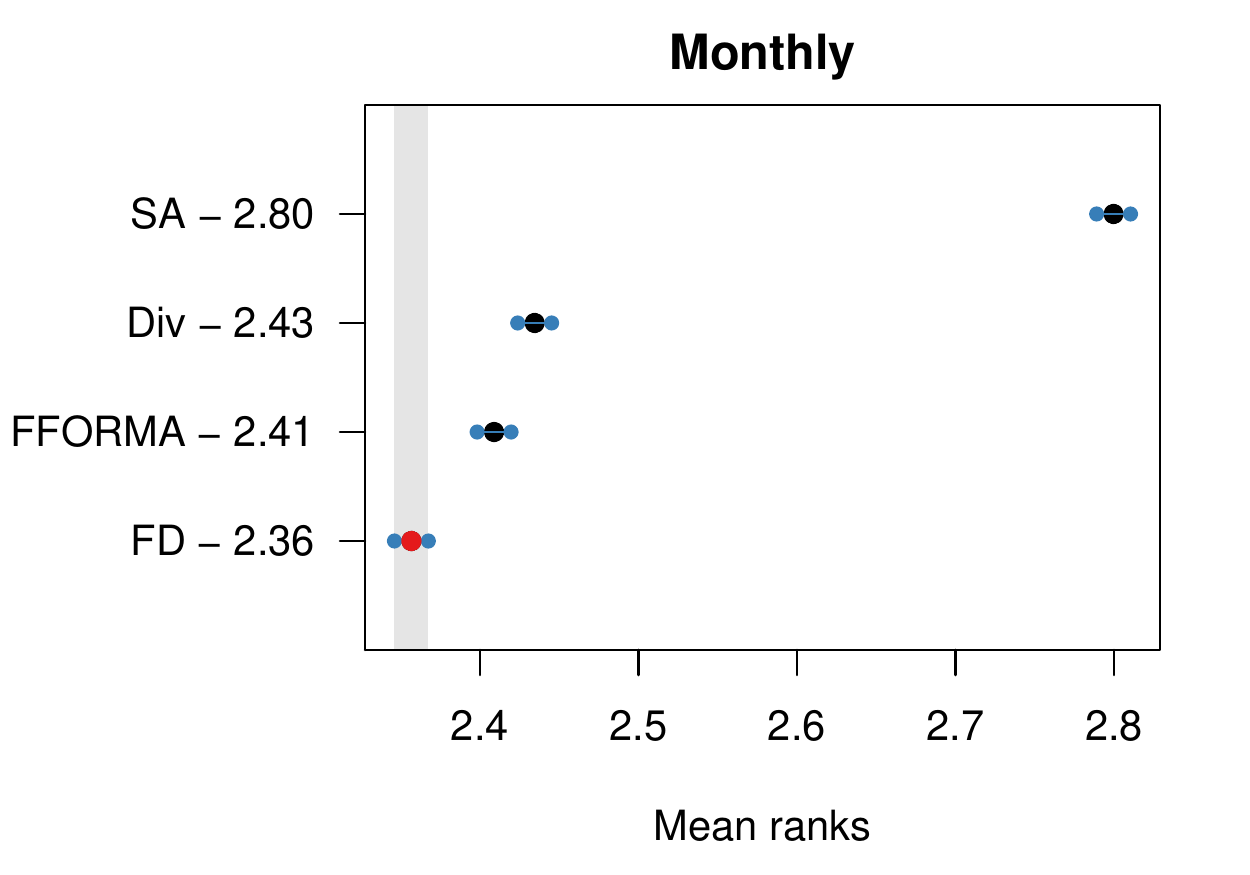}
\end{minipage}
\begin{minipage}{0.45\textwidth}
\includegraphics[width=\linewidth]{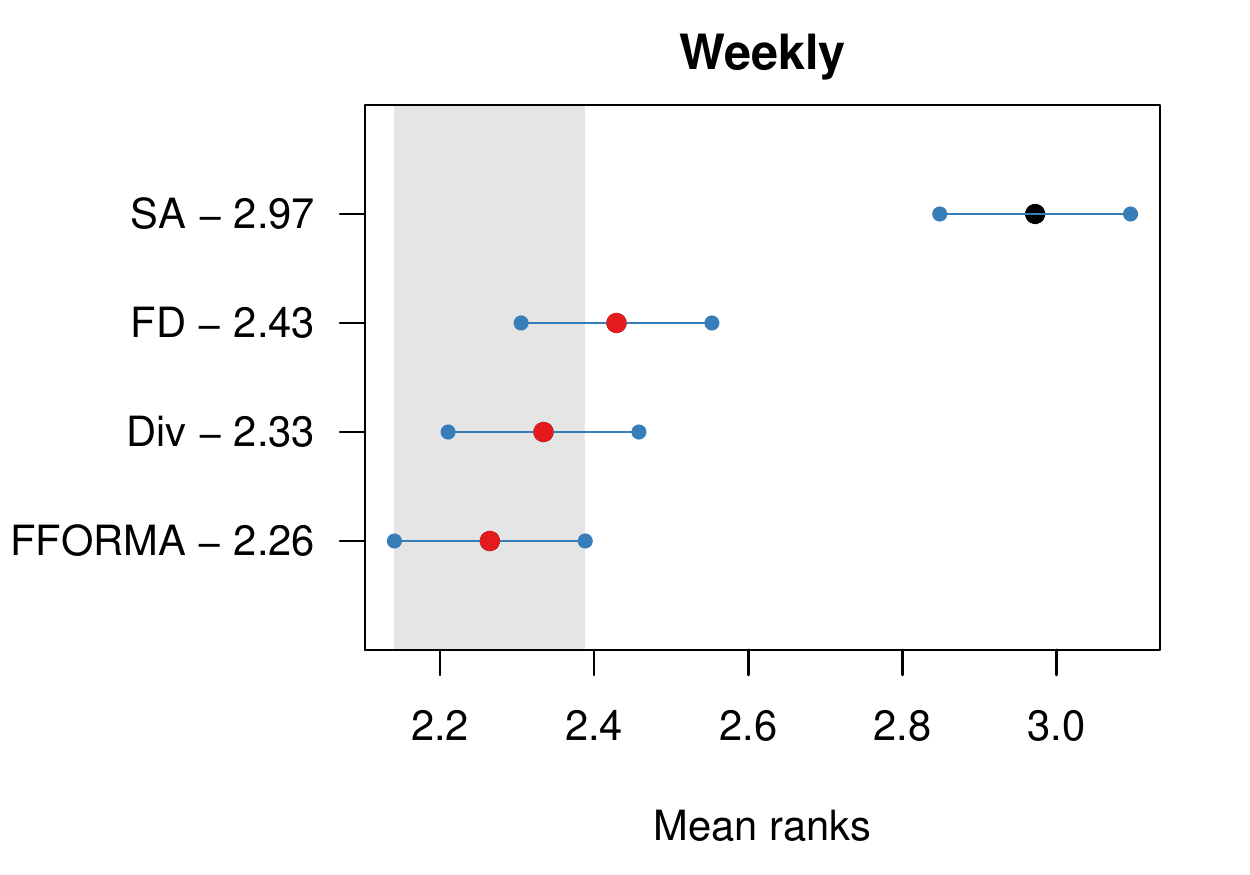}
\end{minipage}
\medskip
\begin{minipage}{0.45\textwidth}
\includegraphics[width=\linewidth]{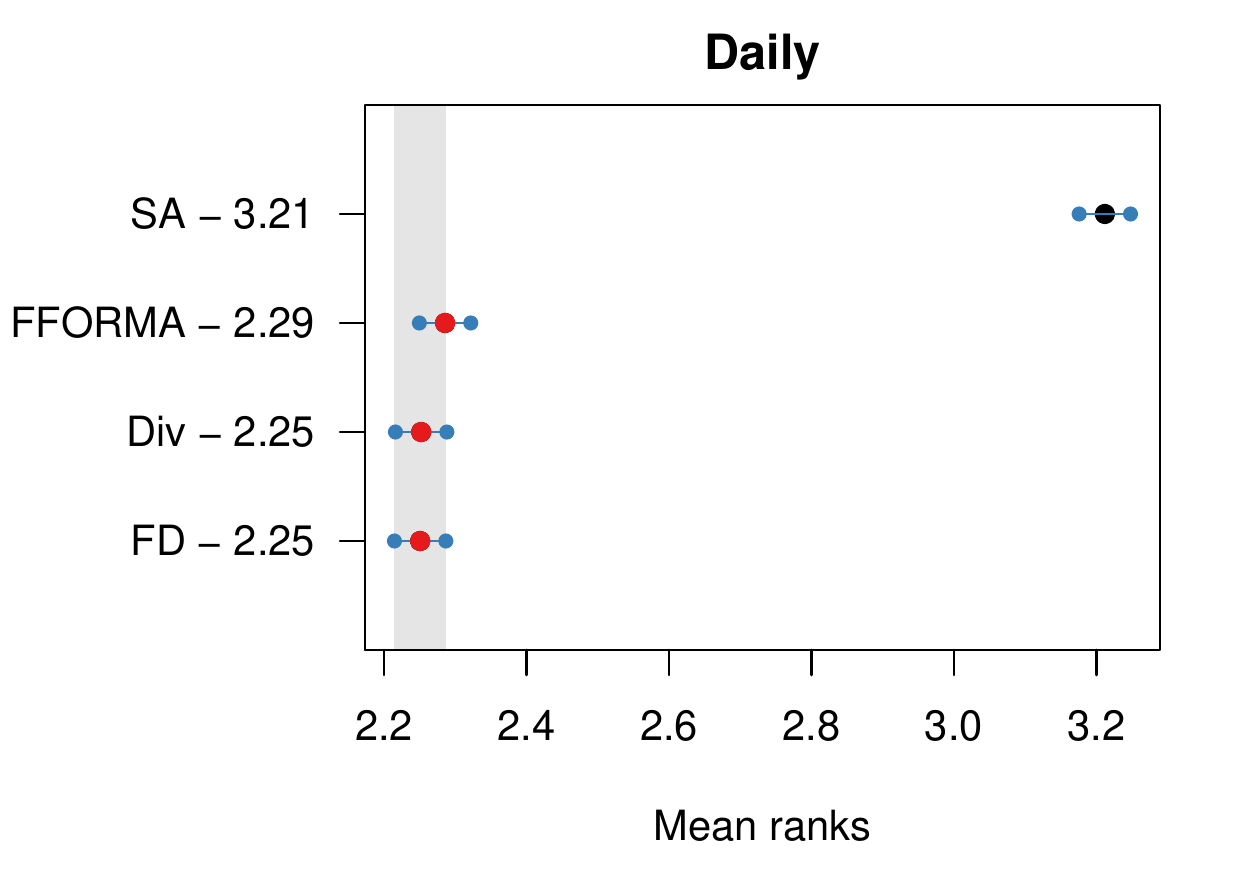}
\end{minipage}
\begin{minipage}{0.45\textwidth}
\includegraphics[width=\linewidth]{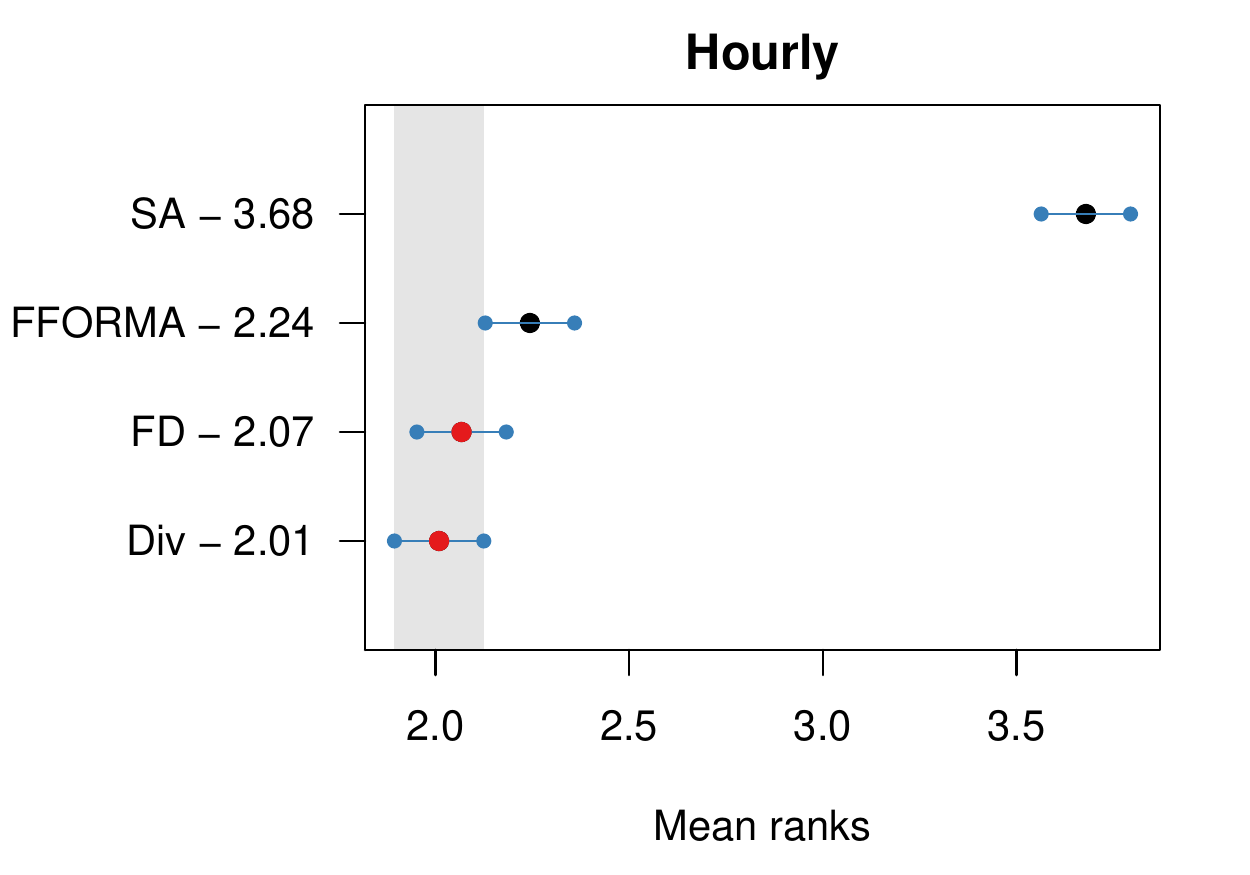}
\end{minipage}
 \caption{MCB tests on the ranks of the MASE errors of \method{SA}, \method{FFORMA}, \method{Diversity} and \method{FD} for each data frequency separately and across all
frequencies (Overall). }
\label{fig:mcb}
\end{figure}

To investigate the statistical significance of the performance differences, we performed Multiple Comparisons from the Best (MCB) test~\citep{koning2005m3} on each data frequency
separately but also over all the M4 series. The aim was to test whether the average ranks of each forecasting method are significantly different from the others. The MCB test was applied
based on the MASE errors as shown in Figure~\ref{fig:mcb}. One can read the results as follows. Lower average ranks are better, althouogh the performance differences between any two
methods are not significant if their confidence intervals overlap.

According to Figure~\ref{fig:mcb}, we observe:
\begin{itemize}
  \tightlist
  \item Although \method{Diversity} outperforms \method{FFORMA} on average, their differences are not significant (see the top panel, ``Overall''). However, \method{Diversity} is more
intuitively appealing and easier to compute.

  \item Overall, \method{FD} performs significantly better than \method{Diversity},  \method{FFORMA}, and \method{SA}. In other words, the diversity features, focusing on
future forecasts, could bring significant improvements to combination forecasts based on traditional time-series statistical features, which only represent information
from time series historical values. That suggests that the two groups of features could complement each other when used for estimating weights for forecast combinations.

  \item Looking at the MCB results for different frequencies, in most cases \method{FD} or \method{Diversity} outperforms \method{FFORMA}. \method{SA} is always significantly outperformed
by the other three approaches.
\end{itemize}

\subsection{Trade-off curves}

We approximate the performance of the proposed \method{Diversity} approach on inventory forecasting by assuming that the prediction intervals produced are directly used
for inventory-related decisions. We produce prediction intervals for various confidence levels: 60\%, 65\%, 70\%, 75\%, 80\%, 85\%, 90\%, 95\% and 99\%. We then consider two quantities. First, we measure the upper coverage level (the percentage of times where the actual outcome is below the corresponding upper prediction interval, i.e., the 97.5\% percentile for prediction intervals produced at a 5\% confidence level). The average upper coverage level is a proxy of the average achieved service level, as it effectively shows the percentage of times that a demanded product was in stock \cite[see also][]{Svetunkov2018-sk} if the prediction intervals of the forecasts were directly used for inventory decisions. Second, we calculate the average upper prediction interval across time series after scaling with the mean value of the historical data for each series. This is a proxy of the holding cost that would be required to achieve the respective service level \cite[see also][]{Svetunkov2018-sk,PetropoulosSiemsenREP}. Rendering the upper prediction interval scale-independent is required as the different series in our data set refer to different quantities (hundreds versus thousands versus tens of thousands units) and we are interested to explore the average effect across all series.

\begin{figure}[!th]
\centering
\includegraphics[width=0.9\linewidth]{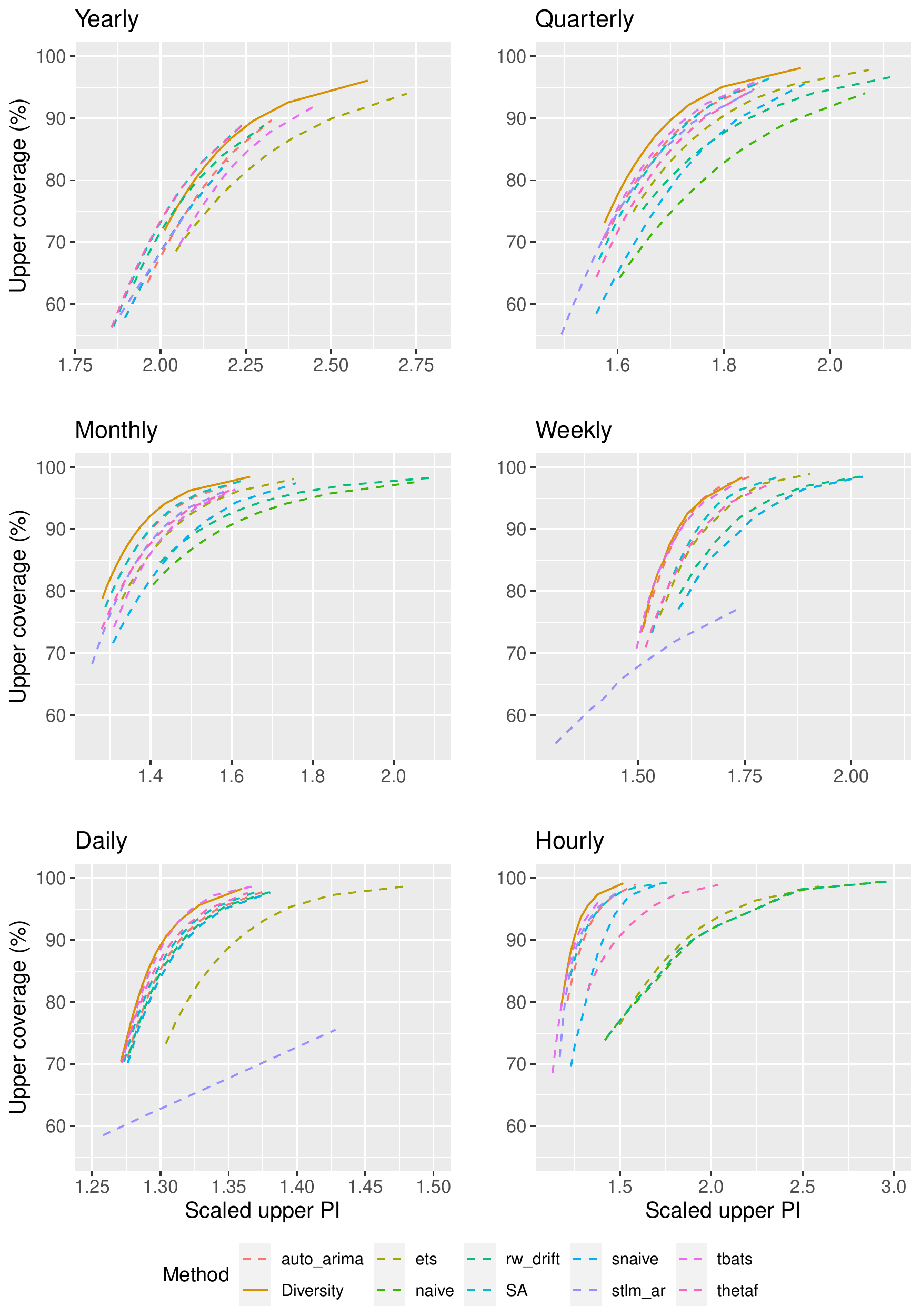}
 \caption{The upper coverage versus the scaled upper prediction intervals across different confidence levels for \method{Diversity}, \method{SA} and the eight individual methods,
for each data frequency separately. \method{Diversity} is shown in solid lines while the other methods are shown in dashed lines.}
\label{fig:ucoverage}
\end{figure}

The trade-off curves for these two quantities (scaled upper prediction interval versus upper coverage), for \method{Diversity}, \method{SA} and the eight individual forecasting methods
in the pool, are presented in Figure~\ref{fig:ucoverage}. The results for each data frequency are presented in a different panel. One can read these graphs as follows \citep{Petropoulos2019-bu}. Assuming a vertical line (similar upper prediction interval, i.e., holding cost), the methods that achieve higher upper coverage levels (higher achieved service levels) should be preferred. Similarly, assuming a horizontal straight line (similar upper coverage levels), the methods with lower scaled upper prediction intervals (lower costs) are better. It can be seen that the \method{Diversity} approach offers a very competitive trade-off between the two quantities considered that outperfroms all other methods for all frequencies. The only exception is the yearly frequency where Theta method performs well in terms of upper prediction interval values but cannot reach high upper coverage levels.

\section{Case study: forecasting fast moving consumer goods}

To highlight the usefulness of diversity in practice, we considered the forecasting of sales of fast-moving consumer goods (FMCG) from a major North American food manufacturer. We focused on
the sales of stock keeping units (SKUs) in two countries, the USA and Canada. The sales are recorded according to monthly frequency and consist of 51 periods, from April 2013 to June 2017.
We set the forecast horizon to 12 and split each series into a training (27 periods), validation, and test sets (12 periods each). Some time series started with zero sales values, which
we trimmed. In those cases in which this trimming process resulted in a training set of fewer than two full seasonal cycles (less that 24 months),
that time series was dropped from the set because it would not be feasible to use to estimate any seasonal patterns. The above process resulted in 955 unique combinations of
SKU $\times$ location.

The model training part of our framework (see also Figure \ref{fig:frame}) was completed using the forecasts corresponding to the observations in the validation set (observations 28 to 39).
The method pool, cost function, and the diversity extraction procedure used in the training process were the same as those used in Section~\ref{sec:experiments}. Once we estimated
the combination model based on diversity, we then applied it to the forecasts for the test set (periods 40 to 51), where we also used MASE and MSIS to measure the out-of-sample
performance. The same approach was used for the \textsf{FFORMA} and \textsf{FD} approaches, similar to the results in Section \ref{subsec:xgboost-results}.

The results from our case study are presented in Table \ref{tab:casestudy}. Much like our main empirical results, we observed that the proposed combination approaches
based on the diversity of the forecasts alone (\textsf{Diversity}) or on  diversity combined with other times series features (\textsf{FD}) outperformed \textsf{FFORMA} and \textsf{SA},
in both terms of point forecast accuracy and estimation of the prediction intervals. It is important to highlight that the results of this case study were based solely on the
955 FMCG series for both the model training and the forecasting phases of our proposed framework (see also Algorithm \ref{alg:A}), thus showcasing that \textsf{Diversity} does not
require massive reference data sets to estimate a diversity-based combination model.

\begin{table}[ht!]
  \centering
  \caption{Comparison of the mean MASE and MSIS values from our diversity-based forecast combination method
    (\textsf{Diversity}), \textsf{FFORMA} and forecast combination with simple averaging (\textsf{SA}) on the FMCG data.  Entries in bold highlight that our method outperforms the
\textsf{FFORMA} approach.}
  \label{tab:casestudy}
  \begin{tabular}{lcccc}
    \toprule
    Method & \textsf{SA} & \textsf{FFORMA} & \textsf{Diversity} & \textsf{FD} \\
  \midrule
  MASE & 0.9555 & 0.9599 & \textbf{0.9365} & \textbf{0.9367} \\
  MSIS & 8.5085 & 8.1254 & \textbf{8.0066} & \textbf{7.9189} \\
    \bottomrule
  \end{tabular}
\end{table}

\section{Discussion}
\label{sec:discussion}

Feature-based forecast model selection and combinations face the challenge of selecting an
appropriate set of time series features that vary according to different domains and
forecasters \citep{fulcher2014highly,kang2019fuma,kang2019gratis}. More importantly,
features' estimation is unreliable when historical data is limited (e.g., for fast-moving
products), or even unavailable when there is no history at all (e.g., for new
products). This study proposes to forecast with forecasts without manually choosing time
series features, yielding comparable performance with top contestants in the M4
competition data with regard to both point forecasts and prediction intervals.

The performance of forecast combinations is potentially highly related to the degree of diversity among the individual forecasts \citep{Armstrong2001-lb,thomson2019combining}.
This study explores how to further improve forecast combinations and attempts to tackle the forecast combination puzzle by exploiting the forecast diversity in the forecasting method pool.
The proposed diversity-based forecast combination can automatically control the combination via measuring the pairwise diversity between forecasts from different sources and linking them,
via a meta-learner, to the accuracy of a test set. If the pool of available forecasts has low diversity, then our approach will approximate an equal-weight combination approach,
which is an appropriate strategy with a lack of additional information. However, if the diversity across the candidate forecasts is high, then the combination weights will be modeled
using two important factors--diversity and accuracy--in arriving at efficient ensembles  \citep{Lichtendahl2020-bj}.

Our empirical results show that the diversity information among individual forecasts used for combination is informative in allocating the combination weights. In fact, using
forecast diversity as the sole time-series feature results in performance that is equivalent to using an array of time-series features calculated on  historical (in-sample) data.
Our approach is not only faster to compute, but also simpler and more straightforward because it does not involve decisions related to which features to include and how to compute them.
As such, it is in-line with the simplicity argument of \cite{Green2015-mu}. In addition, forecast diversity can be used in conjunction with other established time-series features
to boost  forecasting performance.

Research on diversity-based regression/classifier ensembles in the machine learning literature is in line with our findings. Specifically, \citet{liu1999ensemble} used negative
correlation learning to create an ensemble with negatively correlated networks and encourage their specialization and cooperation. \citet{kuncheva2003measures} improve the ensemble
accuracy by measuring the diversity in classifier ensembles. \citet{mendes2012ensemble} reviewed the ensemble approaches for regression and emphasized the importance of regression
diversity. Our study aligns with this line of research in the sense that it aims to improve forecast ensembles by exploiting forecast diversity in the ensemble.

The good performance of our proposition is a result of two factors. First, we built on the
rich and established literature on forecast combinations and explicitly took into account
one of the critical elements in building effective forecast combinations: the diversity of
the forecasts. Second, and in line with the arguments made in the study of
\cite{PetropoulosSiemsenREP}, we explicitly considered the output of the forecasting
models, i.e., the out-of-sample forecasts, rather than simply focusing on information that
relates to the features and characteristics of the in-sample data \citep[like
in][]{montero2018fforma} or how well forecasting models fit the in-sample data. The use of
the out-of-sample forecasts toward making forecast-related decisions is crucial. In
\cite{PetropoulosSiemsenREP}, the evaluation of the representatives of the out-of-sample
forecasts to the actual situation enabled the acceptance or rejection of some forecasting
models (i.e., judging models by their outputs). In our context, out-of-sample forecasts
informed our algorithmic calculations for obtaining weights for forecast combinations by
amplifying the diversity of the final pool of methods being combined.

Another advantage of the proposed approach is its nonreliance on specific families of models. In this study, we focused on linear statistical methods. However, nonlinear methods
could be part of the pool of models in other contexts. Moreover, our approach can  be applied equally to both statistical and judgemental forecasts or even to a combination of the two.
Although in this study we focused on exploring the benefits of forecast diversity in the context of statistically produced forecasts, our approach could be extended toward combining
forecasts from experts, where prior research suggests that the performance of equal weights is hard to beat \citep{Genre2013-ka}. Our suggestion to test diversity-based forecast
combinations also extends the research of \cite{Grushka-Cockayne2017-cy}, who showed that combinations based on trimmed means (excluding the top and bottom $x\%$ forecasts) could
improve the accuracy of overfitted and overconfident forecasts.

Although the proposed method improves point and interval forecasting, a possible future research path is to extend it to probability density forecasting. Meta-learning can presumably
be used to produce a weighted mixture of the forecast distributions from multiple models or to generate a weighted average of the forecast distributions. Diversity could be measured
based on probability distribution distances (e.g., Kullback–Liebler divergence). The loss function in meta-learning could also be adapted to density forecasting principles
(e.g., calibration and sharpness).

One limitation of the current study is that it uses all the available individual forecasts without pooling the most heterogeneous forecasts. Although, naturally, some combination weights
will be close to zero, excluding per series models with poor performance could further improve the forecasting performance. Future research could examine how diversity-based forecast
combinations can be extended to the whole spectrum of selection-pooling-combinations \citep{cang2014combination,kourentzes2018-oa}. Besides, the impact of adding, removing,
and selecting different methods in the approach is also a valuable research topic.

Another limitation of our study is that our empirical results on the M4 data are, in
theory, not directly comparable with any of the original contestants of the M4
competition.  Access to the postsample data allows for experimentation, testing, and
hyper-parametrization, none of which were available to the original participants. Although
we do not directly use future data values to inform our forecasts, we still cannot claim
that our approach would have performed better than \method{FFORMA} in the M4 competition
because our approach was never submitted. Even if we had also applied the proposed
approach to a real application, the same forecasting method pool is used with most of the
methods being drawn from results of societal activities and being used with linear
approaches of various types. As such, a final avenue for future research would be to
compare our framework against the performance of other established benchmarks in more/new
data sets, including data with nonlinear and intermittent patterns.

\section{Conclusion}
\label{sec:conclusion}

In this paper, we proposed to use forecasts to improve forecasting performance. In essence, we
measure the pairwise diversity among the forecasts from the methods in a pool for each time
series and use these measurements as a group of features linked with the out-of-sample forecasting performances
of the individual forecasting methods. We show that our approach--simply using forecast diversity--achieves equivalent performance
to the use of manually selected time series features calculated from historical data.  Combining diversity and other statistical features
(depicting future and historical information, respectively) can be further advantageous.

Our approach provides an automatic and flexible tool for forecasting practice. Our proposed approach has the following merits. First, forecasters do not need to tackle
the issue of feature selection when carrying out feature-based forecasting. The calculation of
diversity is straightforward, easy to implement, and interpretable. Second, forecasts from
any method (including statistical methods, nonlinear techniques, and judgment) can be easily used
within our approach. Finally, our proposition offers improved point forecast accuracy coupled with better performance for interval forecasts.

\section*{Acknowledgements}

The authors are grateful to the editors and three anonymous reviewers for helpful comments
that improved the contents of the paper. Yanfei Kang is supported the National Natural
Science Foundation of China (No. 72171011, and No. 72021001) and the National Key Research
and Development Program (No. 2019YFB1404600) and Feng Li is supported by the Emerging
Interdisciplinary Project of CUFE and the Beijing Universities Advanced Disciplines
Initiative (No. GJJ2019163). This research is supported by the high-performance computing
(HPC) resources at Beihang University.

\bibliographystyle{agsm}
\bibliography{mybib}
\end{document}